 \documentstyle[NATO,numreferences]{crckapb}

\begin{opening}
\title{Energy Flow in the Universe} 
\author{Craig J. Hogan}
\institute{Astronomy and Physics Departments \\
 University of Washington\\
 Seattle, Washington 98195, USA}
\end{opening}
 \runningtitle{THE CRCKAPB STYLE FILE}
\begin{document}
\begin{abstract}
 A  brief but broad survey is presented of the flows,
forms and large-scale transformations of mass-energy  in
the universe, spanning a range of  about twenty orders of magnitude 
($\approx m_{Planck}/m_{proton}$) in space, time and mass. Forms of energy considered
include electromagnetic radiation, magnetic fields, cosmic rays, 
gravitational energy and gravitational radiation, baryonic matter, 
dark matter, vacuum energy, and neutrinos; sources considered include
vacuum energy and cosmic expansion, fluctuations and gravitational collapse, 
AGN and quasars, stars, supernovae and gamma ray bursts. 

\end{abstract}

\section{Global Energy}
 
Everything that happens is a transformation of mass-energy.
Starting with inflation and the Big Bang, mass-energy flows
through  and organizes structures spanning an enormous range
of scales--- lengths and 
times  from billions of years down to milliseconds.
These notes survey the  main features of   
cosmic energy cycles on a global scale and trace  the 
causal links between them. 

An absolute    luminosity limit for anything is imposed by   General
Relativity.
Suppose a
 sphere of radius $R$ is filled 
with light of total mass-energy $Mc^2$
 and
released an an instant; the energy has left the sphere after a time
$R/c$, with an average luminosity of $Mc^3/R$. But the gravity of the
energy imposes a limit on how small $R$ can be: if it is smaller
than the Schwarzschild radius  $R_S=2GM/c^2$, no light can 
escape at all since it is within the event horizon of a black hole.
The maximum luminosity of any source is therefore 
\begin{equation}
L_{GR}=c^5/2G=
m_{Planck}^2/2=1.81\times 10^{59}{\rm erg \ sec^{-1}},
\end{equation}
 independent
of mass. (We have expressed $G$ in terms of 
  the   Planck mass  $m_{Planck}=\sqrt{\hbar c/G}=1.2\times
10^{19}$ GeV $\approx 10^{-5}$g,
according to the convention $\hbar=c=1$. It is tempting call $L_{GR}$ the ``Planck
Luminosity'' since it corresponds to a Planck mass per Planck time,
but in fact Planck's constant $\hbar$ cancels out when using
units of luminosity--- $L_{GR}$ does
not depend on quantum mechanics.)
This 
luminosity represents an upper bound on the rate of
energy transformation of any kind, on any scale.

For objects within the universe, nothing approaching this luminosity
has ever been observed--- not because of size, but because
  radiation interacts with matter (indeed, it must interact 
to be generated) and therefore takes time to
``leak out'' of a system. The maximal limiting luminosity  
requires both an efficiency   close to unity and an interaction time  
close to a light-travel time.  Even neutrinos interact more strongly
than this in dense  collapsing cores of stars where they are
copiously produced.
 The only situation    where the   limit
seems likely to be approached is in   production of gravitational
radiation from merging black holes of comparable mass,  
  events which may eventually
be observable in gravitational waves.\cite{gw,gw3}

Roughly speaking, the Big Bang 
 saturated the absolute  bound $L_{GR}$.
 At any time 
 during the early radiation-dominated
phase of the early universe, this is about equal
to the energy of cosmic radiation 
within a Hubble volume divided by a Hubble time. 
If the universe today is dominated by vacuum energy
(a cosmological constant or some other form with $\Omega_\Lambda\approx
0.7$)\cite{snesearch1,snesearch2}, then the  ``$PdV$ work'' being done right
now in each  Hubble volume is  also comparable to $L_{GR}$. (A comparable rate
of  transformation occurred  during inflation
and during reheating.) The cosmic background radiation is less
than this today by a factor of about $10^4$ because of redshifting.

The comparable amount of energy 
locked up as rest-energy of Dark Matter has not substantially 
interacted with anything microscopically
 for a long time--- for most candidates,
at least since the weak interactions decoupled. 
However, a significant flow of energy occurs 
in the dark matter via gravitational collapse.
This energy 
predates even the cosmic radiation, originating in   primordial fluctuations 
in  binding energy which produce  cosmic structure, which probably date 
back to inflation. These perturbations are injecting
observable energy flows into the universe today, into
the nonvacuum components--- the Dark Matter, with
$\Omega_m\approx 0.3$, and the baryons, with $\Omega_b\approx 0.03$. 
Their dimensionless amplitude is about $10^{-5}$--- this is
the fraction of total mass-energy available as free energy 
in this form.
There are of course small flows caused by radiation
 temperature anisotropies but
the dominant effect   is gravitational collapse, which in turn 
creates kinetic motion in the dark matter and causes  heating
of baryonic  gas by compression and shocking. This process  heats the bulk of
cosmic baryonic matter to temperatures   of about
 $10^{-6}m_{proton}\approx 10^7$ K, creating a
pressure sufficient to keep most of the baryons  from falling into
galaxies and stars.\cite{gravheat,baryons} The gas achieves a  steady state at
this characteristic temperature, where the ``hierarchical heating''
is balanced by adiabatic losses to the expansion (and thereby slightly
modify the global expansion rate.)

 Roughly speaking, the cosmic
fluctuation amplitude of $Q\approx 10^{-5}$ injects $10^{-5}\rho c^2 $ 
of energy  per Hubble time in the cosmic web, with typical velocity scale
$Q^{1/2}\approx 10^{-2.5}c$ and typical size $Q^{1/2}/H$
corresponding today to galaxy superclusters. The power in
this form of heating is   substantial, about $10^{52}$ erg/sec
in the baryons and $10^{53.5}$ erg/sec in the dark matter. 
Because it radiates inefficiently
this dominant reservoir of intergalactic matter is practically
invisible except as a diffuse soft X-ray  background.

\section{Stars}

The other forms of energy involve   the small  fraction of baryons
which make it into galaxies.
The total density of baryons in galaxies, including
all stars and their remnants as well as star-forming gas,
adds up to  less a quarter of the baryons, or less than
one percent of the total density. A  small fraction of this
material makes it into AGN and supernovae. Somehow the activity is coupled
so that these things all contribute roughly comparable total energy
budgets.

Most light since the Big Bang has been made by ordinary main-sequence
 stars. They are still forming  from gas, although most
of the stars in our past light cone formed about $10^{10}$
years ago---  about
$10^{10}$ of them in each of $10^{10}$ galaxies in our reference volume,
formed over roughly $10^{10}$ years. Each one lasts for a long time;
most of the mass-energy budget is in low mass stars which last
for billions of years ($10^{10}$y for 1 $M_\odot$).  The total power in stars 
in all galaxies over all of  cosmic history 
is now close to being accounted for in cosmic backgrounds from
the optical to the far infrared.\cite{optical,fir,dirbe,metals}
 The distribution in time
is estimated   from redshifts of 
 directly imaged galaxies\cite{madau} and the backgrounds are 
now close to being resolved, so the spacetime distribution of
the source populations are close to accounted for. The total
flux  is about 1/30 of the cosmic background radiation, approximately
equally distributed between direct light from stars and 
reradiated light from obscuring dust, and mostly
radiated since a redshift of about two, when the universe
was three times smaller than today. 

This energy ultimately
derives mostly from the conversion of hydrogen to helium,
with some contribution from synthesis to heavier elements.
The enrichment history is  recorded as fossil abundances in stars
and can also be traced directly in high-redshift absorption
 systems\cite{absmetals}.
The total amount of light agrees with   the total  production of  
elements by  stars\cite{metals} if  the   metals
are mostly ejected from the stellar parts of galaxies and join 
the dominant intergalactic gas. This large-scale sharing of metals 
 is confirmed from X-ray line emission in galaxy clusters\cite{clustermetals}.

Stellar formation as well as stellar instabilities
(important especially at the end of the stable nuclear 
burning stage) all occur at roughly the
Chandrasekhar  mass, corresponding to  a number of protons  
\begin{equation}
N_C= 3.1(Z/A)^2 (m_{Planck}/m_{proton})^{3}\approx 10^{57}
\end{equation}
(shown here with its classical definition as 
the limiting mass of an electron-degeneracy-supported dwarf; $Z$ and $A$ are the
average charge and mass of the ions, typically $Z/A\approx 0.5$ and
$M_C=1.4M_\odot$, where $M_\odot= 1.988\times 10^{33} g\approx
 0.5 M_*$ is the mass of the Sun.) 
One way or another 
the large numbers in Table 1 for baryonic flows all ultimately derive from 
the large number $(m_{Planck}/m_{proton})\approx 10^{19}$;
this is true even for the global cosmological   quantities,
since the age of the universe now is about the lifetime of a 
star.

\section{  Quasars}

In the centers of galaxies, a small fraction  of material 
(up to   about $10^{-2}$ of the stellar mass, so on the order
of $10^{-4}$ of the total  mass-energy)
 accumulates and organizes
itself into a different kind of engine, called
quasars,  active galactic nuclei or AGN,
which generate their enormous
 power  from gas interacting with
massive black holes.  Although events can occur in quasars quickly
(on timescales of days or less, the Schwarzchild time
for the massive holes) the bright phase lasts for
tens of millions of years, determined by the behavior of the surrounding
gas. The bright activity of quasars peaked at redshifts
of about two and is much less today\cite{qlf},
but the black hole remnants of $10^6$ to $10^9M_\odot$ reside still in the centers
of most galaxies including our own.\cite{galbh} The total number $N$ of
  quasars is thus about the same as the number of galaxies,
with a tendency for large bright ones to lie in the biggest galaxies.
The overall
energy from quasars is not much less than that from stars, due to
their large efficiency in converting rest-mass into energy.
They dominate the energy budget of the universe for most hard radiation
such as X-rays and gamma rays
(with some competition from supernovae), and are mainly responsible for ionizing
the intergalactic gas.\cite{igm}

Active nuclei   derive all of  their
electromagnetic  energy from gravity--- either  from
the binding energy of the infalling material, or from the rotational
energy of the black hole.\cite{engine,engine2,engine3} 
Gas falls in and forms a disk near the hole,
fattened by heating into a torus or corona.  Magnetic fields help to extract the
orbital and spin
energy and also to channel some of it into ``Poynting jets''
 of relativistic matter
moving so close to the speed of light that a particle's kinetic
energy is many times its rest mass, with Lorentz factors of 
$\Gamma=E/mc^2\approx 10$.  Light emerges at all wavelengths,
from radio to gamma rays, reflecting activity on  many scales and
  nonthermal radiative processes involving relativistic
particles, magnetic fields and bulk kinetic energy of matter. 
There should be a comparable luminosity in cosmic rays,
a small portion of which is channeled into high energy neutrinos.

 The accretion rate
of matter and hence the luminosity is approximately regulated
by feedback on the 
gas accretion.
For both AGN sources and massive stars 
 the characteristic luminosity   is the    Eddington limit,
$L_E= 3GMm_p c/2 r_e^2=1.25\times 10^{38}
(M/M_\odot)$ erg/sec (where the
 classical electron radius $r_e= e^2/m_ec^2$), above which radiation
 pressure   outwards on ionized gas
exceeds gravitational attraction; brighter sources tend
to disassemble themselves. An Eddington-limited
source lasts a 
    Salpeter  time  
$Mc^2\epsilon/L_E=4\times 10^8\epsilon$ y, which is independent of
mass but does  
   depend on 
the overall efficiency  $\epsilon$ of extracting rest mass $Mc^2$, 
which may be
as large as tens of percent for material near a black hole.   
Significant variability occurs on all timescales down to  the Schwarzschild
time of the black hole, $R_S/c=  10^{-5} {\rm sec} (M/M_\odot)=
  100 {\rm sec} (M/10^7M_\odot)$.

Because mergers of galaxies are common, it is likely
that mergers of their central holes are common. If  
$10^{10}$ galaxies within the Hubble volume
 each merge about once per Hubble time,
there is about one such event per year in our past light cone, releasing about
$10^{62}$ ergs for a $10^8M_\odot$ hole. Such an event,
with a luminosity of $\approx L_{GR}$,
far outshines  all other sources put together for a 
Schwarzschild time (on the order of  minutes for a $10^8M_\odot$ hole).
Even at an average rate of one per year, the  gravitational wave luminosity of
the universe radiated from these mergers   is on average
comparable to all the other forms of energy combined, stars and everything. 
These waves have not yet been detected, but they
are in principle easily detectable with spaceborne gravitational
wave detectors such as LISA\cite{lisaligo}.

\section{Superstars and Supernovae}

Smaller but equally violent energy releases
 occur as byproducts of
instabilities in dead or dying   stars. When a star exhausts
the nuclear fuel in its core, it is no longer stable; the core collapses
seeking a new equilibrium, and the release of energy from this collapse
blows off the enveloping material. 
The outcome depends  on the mass and composition
of the star. A small star like our Sun
will blow off about half of its mass, the rest of it left behind 
in a   white dwarf, a glowing ember of still-unburned
nuclear fuel (e.g. He, C, N, O, Ne,...), about 10,000 km diameter
(about the size of the Earth) and a million times the density of 
ordinary matter, stabilized by electron degeneracy pressure
against gravity.  More massive
stars create    iron cores  
above the Chandrasekhar limit of $1.4M_\odot$ at
which electron degeneracy support fails, and   collapse  
to a neutron star with a diameter of only  about 10 km
and the same density as  an atomic nucleus.  
Massive cores above a few solar masses cannot
be supported by neutron degeneracy or gluon pressure,    and
 collapse all the way to  black holes.

Collapse of these remnants
releases  gravitational binding energy.  
Smaller and denser objects create more ($\propto 1/r$) and
faster ($\propto \rho^{-1/2}$) energy release.
  White
dwarf formation ejects a planetary nebula at high velocity;
neutron star formation leads to a Type II supernova
explosion.\cite{sn2} 
Other spectacular effects occur when remnants live in binary systems
and perform a whirling dance
with  normal stars or with each other. Accretion onto compact remnants
from companion stars 
leads to cataclysmic X-ray sources,
and accretion onto a white dwarf can  trigger a nearly complete
 nuclear deflagration and disruption, leading to a Type Ia
supernova.\cite{sn1a} 

The scale of energy budgets of cataclysms derives from the mass-energy
of the stellar remnants, with a basic scale set by
  the rest mass of the sun
 $M_\odot c^2= 1.8\times 10^{54}$ ergs.
The nuclear energy available from a white dwarf is about 
$10^{51}$ ergs, most of which is liberated when a Type Ia supernova
explodes, mostly as blast energy.
The binding energy of a neutron star is about
$ 10^{53.5}$ ergs, almost all  of which is radiated as 
neutrinos during a Type II supernova. (These were directly detected
from supernova 1987a; the sum of such events
over the Hubble volume leads to a soon-to-be-detectable neutrino
background\cite{snu1,snu2}.)
 A small fraction of the neutrinos
as well as a bounce shock from the neutron star couple
to the enveloping material, dumping heat which  ejects  it at high velocity.
About $10^{51}$ erg emerges as blast energy, less than ten percent
of this as light. Heat, light, heavy elements, magnetic fields,
kinetic motion
and cosmic rays 
all carry a substantial amount of energy far away and spread over
a volume vastly larger than their source, providing a 
regulatory cycle and coupling of 
energy and material flow.

The overall energy budgets of the supernovae are again surprisingly
close to that of the stars and AGNs, although most of the SN energy budget
is emitted in neutrinos, with a small fraction 
as blast energy and an even smaller fraction as light. Even though
 only about a percent of stellar
mass participates in supernovae, the $\nu$ production
 efficiency is much higher
than nuclear energy ($\approx 0.1$ as opposed to $0.007$).
Although the energy sources for the different types
of supernovae are completely different (and the Type Ia even leaves
no compact remnant), their blast energies are similar.  
Both eject a substantial mass  of 
chemically enriched and freshly-made radioactive material 
which powers a glow lasting for months. 
Also by coincidence, the cosmic rates of the different types of 
supernovae are comparable in spite of the very different progenitors;
SNeIa are just a few times brighter, and a few times rarer, than SNeII.

\section{  Fireballs and Hypernovae}

Long shrouded in mystery, gamma ray bursts  now seem
to 
make use of the same compact remnants and  combine features
of both  supernovae and of quasars. 
Apparently, even these most exotic of sources do not require
new cast--- only new roles, new settings and combinations
for the familiar players, neutron stars and black holes.
They are a kind of naked 
spinning supernova and miniature quasar wrapped into one.

The main event in a  gamma ray burst is
 a ``relativistic fireball.''\cite{fireball1,fireball2}
A burst of much energy in a small space   results in 
an expanding plasma of photons, electrons and positrons,
neutrinos and antineutrinos. 
There is enough scattering for
 matter to behave like a fluid, though few enough baryons
(less than $10^{-4}M_\odot$) not to inhibit acceleration with 
particle rest energy, so the
 relativistic fluid expands  very close to the speed of light,
 with a  Lorentz factor $\Gamma\ge 100$. The kinetic
energy is dissipated in shocks and radiated as gamma rays at
radii of $10^{13}-10^{15}$ cm, the scale of the solar system.
Because of Lorentz beaming   any observer sees at most a small
relativistically-blueshifted
patch of the fireball, allowing rapid variability.
(The fireball may also itself be beamed, visible only from
some directions.)
The interaction with the environment as well as the 
beaming leads to a wide  variety of
events, with variability down to  
milliseconds but a duration up to hundreds of seconds,
and optical  afterglows which remain observably bright for a few days--- as large
a temporal dynamic range as in a quasar, but scaled small.
  The energy of
the brightest gamma ray bursts is typically  estimated in 
both gamma rays and optical afterglow energy to be $10^{53.5}$ erg\cite{energy}
(or in extreme  cases $10^{54.5}$ erg,
assuming isotropic emission; allowing for anisotropic beaming, the total
energy budget could well be   less than this.)

The fireball, like a supernova,
 is created by a cataclysmic combination of
stellar remnants. A favorite current model invokes a stellar-mass black hole
 or neutron star 
surrounded by a torus of neutron-density material---
essentially, a donut-shaped neutron star surrounding a more
massive black hole.\cite{donut,donut2}
 This   donut+hole  system
resembles   a very dense, scaled-down version of the quasars,
with magnetic fields, now in combination with the highly dissipative neutrinos,
extracting energy from the spin of a black hole and/or the orbital 
energy of the torus.  
The Schwarzchild time for a $10M_\odot$ hole is $10^{-4}$ sec
so the smallest timescales   can be explained;
as in quasars the disk and the event last for much longer than
this dynamical time.

 A spinning black hole can liberate
up to $0.29Mc^2$ or $5\times 10^{54}$ ergs for a 10$M_\odot$ hole;
material in a disk can liberate up to $0.42Mc^2$ or $10^{53}$ ergs
for a 1$M_\odot$ disk. There is thus ample energy
in a stellar-mass ``microquasar''   to power the high-$\Gamma$ burst of  gamma rays and
a lower-$\Gamma$  optical afterglow.
It seems likely that beaming  should often produce  the latter without
the former--- a new population of objects which would lack
 gamma rays,  perhaps  
appearing  like short-lived, very bright  SNeII. These may
have already been noticed in distant supernova surveys
\cite{snesearch1,snesearch2}
 but in any case are not
common (at most about $10^3$ per day) so they are likely not important in
the overall energy budget.

It is not clear exactly how this   configuration is produced, 
but several ideas fit well into stellar evolution scenarios.
One model is a ``hypernova''\cite{hypernova}, which is like a particularly massive
Type II supernova
core collapse but with a collapse of some of the material inhibited
or delayed by rotation. The middle forms a black hole, some of the rest  
forms the neutron torus. Alternatively, two neutron stars in
a close binary (themselves formed from earlier supernova explosions)
might eventually coalesce by gravitational radiation of their
orbital energy; the mass can exceed the maximum mass
of a stable neutron star, leading to a black hole surrounded
by a dense neutron torus.\cite{binaryns} Either of these 
 scenarios plausibly
leads to the enormous magnetic fields required to form a 
 microquasar.  One way to distinguish them observationally
is by observing where the bursts occur: the first picture produces
a burst after only tens of millions of years, while the second
may be after billions of years, and should produce bursts far
from star-formation regions.\cite{hypernova}
\begin{table}[htb]
\begin{center}
\caption{Scales of Time, Energy, and Power in the Universe}
\begin{tabular}{llllllll}
\hline  
Energy & Luminosity  & Duration & Rate $R$
 & $N_{active}$ & Energy &Power 
 & total $N$
 
\cr
source& $L$/ object &  $D$/ flash & $=NH$&$=NHD$&$E=DL$&$ I_{tot}=RE$&
in
$V_0/H_0$
\cr
{}& (erg/sec) & {} & {}&{}&(erg)&(erg/sec)& $=R/H$ 
\cr 
\hline

Big Bang& $\approx 10^{59}$ &$10^{12}$sec& 1/$10^{12}$sec    &1&
$10^{71}$ &$10^{59} $ & 1\cr

$\Lambda$& $10^{59}$ &$10^{10}$ y&1/$10^{10}$ y &1&
$10^{76}$ &$10^{59} $ & 1\cr

$Q=10^{-5}$& $10^{46}$ &$10^{10}$ y&1/$10^{2.5}$y
  &$10^{7.5}$&
$10^{65}$ &$10^{54} $ & $10^{7.5}$\cr

Stars ($1M_\odot$)& $10^{33.5}$ &$10^{10}$ y&$10^{10}$/y&$10^{20}$&
$10^{51}$ &$10^{53.5}$ & $10^{20}$\cr

AGN($10^7 M_\odot$)& $10^{45}$ &$10^{15}$ sec&1/y&$10^{7.5}$&
$10^{60}$ &$10^{52.5}$ & $10^{10}$\cr

AGN($10^9 M_\odot$)& $10^{47}$ &$10^{15}$ sec&0.01/y&$10^{5.5}$&
$10^{62}$ &$10^{52.5}$ & $10^{8}$\cr

GW($10^8 M_\odot$)& $10^{59}$ &$1000$ sec&1/y&$10^{-4.5}$&
$10^{62}$ &$10^{54.5}$ & $10^{10}$\cr

SNeII($\nu$)& $10^{53}$ &seconds&1/sec& 1&
$10^{53}$ &$10^{53 }$ & $10^{17.5}$\cr

SNe(O/IR)& $10^{43 }$ &$10^{6.5}$ sec&1/sec&$10^{6.5}$&
$10^{49.5}$ &$10^{49.5 }$ & $10^{17.5}$\cr

GRB($\gamma$)& $10^{53}$ &seconds&1/day&$10^{-5}$&
$10^{53}$ &$10^{48}$ & $10^{12.5}$\cr

 GRB(O/IR)& $10^{48}$ &$10^{5}$sec&1/day&1&
$10^{53}$ &$10^{48}$ & $10^{12.5}$\cr
\hline
\end{tabular}
\end{center}
\end{table}

\section{Energy Budgets}
Table 1 shows a broad summary of energy flows of various kinds:
the cosmic microwave background; the accelerating universe;
the free energy injected by cosmological fluctuations and gravitational
instability; normal stars in galaxies; small and large active galactic nuclei;
gravitational waves from mergers of AGN engines in galaxy mergers;
neutrinos from core-collapse supernovae; optical radiation 
and blast energy from Type I and Type II supernovae; gamma
ray and optical emission from gamma ray bursts. 
Except for the Big Bang (which refers to the radiation-dominated
epoch of the universe), 
entries in the table refer to events at moderate redshift
(less than a few)
 out to distances of the order of the 
Hubble distance $cH_0^{-1}$ or about 14 billion light-years
for a Hubble constant $H_0=70 {\rm km\ sec^{-1}\ Mpc^{-1}}$,
in a reference ``Hubble Volume''
$ V_0\equiv 
4\pi c^3H_0^{-3}/3=3\times 10^{11}{\rm Mpc}^3$,
containing about $10^{9.5}$ giant galaxies (with a luminosity
$\approx 2\times 10^{10}L_\odot$
each),
and a spacetime volume $V_0/H_0$.
 Within this spacetime volume there have been about $N$ events
 of each kind, from one Big Bang to $10^{20}$ stars.
The numbers represent  
order-of-magnitude averages over moderate redshifts
$0\le z \le 3$, which includes the bright epochs of star
formation and quasars.
The entries  show  the luminosity $L$ of each single event;
the   typical
 duration $D$ of each event; the  
 rate $R$ at which new events appear; the 
number $N_{active}$ of events active at any given
time; the energy $E$ released in the designated form by
each object;  the power $I_{tot}$ produced by
each population in the entire Hubble volume; and 
the total number $N$ in the Hubble spacetime volume. 
The ubiquitous appearance of numbers like $\approx 10^{20}$ can be traced in all
cases to $m_{Planck}/m_{proton}\approx 10^{19}$.
 Not shown are the timescales of most
 rapid variation; for each source this
has a dynamic range of order $(m_{Planck}/m_{proton})^{1/2}$ , extending for
quasars down to less than
 a day and for compact sources down to small fractions of a second.
 Recall that one day=$10^5$ sec, one month=$10^{6.5}$ sec, one 
 year=$10^{7.5}$ sec.
For AGN and GRB, the energy budgets in kinetic energy,
Poynting flux, and cosmic rays are comparable to the nonthermal
electromagnetic budgets shown; for supernovae, these forms
are somewhat less.
The largest energy by far is the work being done by the cosmological
constant negative pressure in creating new vacuum energy, which
replaces the bulk of the entire mass-energy content of the universe
in a Hubble time.

\section{Cosmic  Ecology}

A glance at $I_{tot}$  in table 1
shows a remarkable coincidence:
  in spite of 20 orders of magnitude variation in
mass and timescale (and $N$),
the integrated power is comparable for all of these populations
if we count the GRB's as a subclass of supernovae.
 This coincidence   can be understood if there 
are feedback loops  controlling the release of energy--- 
 a globally regulated choreography coupling
the 
formation rate of stars, the events leading to their
death and the transformation and ejection of the elements,
the formation of galaxies and
quasars and the feeding of their central engines.

This may be just a coincidence, or it may be a hint that
stars,  galaxies and 
indeed the  universe behave as ``whole systems'' controlled by 
nonlocal interactions between
interdependent parts spanning a large range of scales.
Many mechanisms are available to provide the coupling:
radiation, magnetic fields\cite{mag}, cosmic rays, and fast fluid
flows. 
Although the scales of the individual sources all derive directly
from fundamental
physics, their frequency in the cosmos depends on this
poorly understood
``cosmo-ecology'' of interacting systems. 

Another point is the sheer dynamism of the sky on all timescales.
The rate $R$ of many new events include a range, from seconds to years,
accessible to direct surveys. Somewhere in the sky
a new observable supernova appears every second, with over
a million brightly shining at any time; on average
a new quasar appears every year, with tens of millions shining
at any time.   The dynamic range of variability
includes a range, from milliseconds (for GRB's)  to years (for quasars),
 accessible to
direct monitoring. Astronomical and data exploration techniques
have hardly started to sample what is happening. 

Have we seen it all, thought of everything that could happen,
already explored the entire range of things that could be
happening out  there? These
questions hang in the air whenever new experiments are contemplated,
and for some large projects, such as  gravitational
wave  detectors, are major strategic
concerns.\cite{gw,gw3,lisaligo,gwprojects,gw2}  As the experience with gamma
ray bursts shows, for even the most exotic   sources the  ancient optical band
still holds vital information for uncovering the physics
of the sources. 
There are almost certainly new combinations of familiar
players (such as flares of stars being eaten by dead quasar black
holes) which exist but are not yet found. Microlensing\cite{macho}
 and supernova surveys\cite{snesearch1,snesearch2}
 have shown what
CCD arrays and data-mining can do; these technologies promise
  to
expand  the scope, depth and  precision of the digital exploration
of the time domain by orders of magnitude in the next few years
and reveal a still richer phenomenology.\cite{projects}
  
\acknowledgements

This work was supported at the University of Washington
by NSF and NASA, and at the Max-Planck-Institute f\"ur
Astrophysik by a Humboldt Research Award. I am grateful for the 
hospitality  of the Isaac Newton Institute for Mathematical
Sciences, Cambridge.



\begin{thebibliography}{}

\bibitem[1]{gw}Haehnelt, M. G., Mon. Not. Roy. Astr. Soc. 269, 199-208 (1994)
\bibitem[2]{gw3}Nakamura, T., Sasaki, M., Tanaka, T.,\& Thorne,
K. S., Astrophys. J. 487, L139-L142 (1997) 
\bibitem[3]{snesearch1}Riess, A. G. et al., Astron.J. 116, 1009-1038 (1998) 
\bibitem[4]{snesearch2}Perlmutter, S., Astrophys. J. 517, 565 (1999)
\bibitem[5]{gravheat}Cen, R. \& Ostriker, J. P., Astrophys. J.,
in press (astro-ph/9806281) (1999)
\bibitem[6]{baryons} Fukugita, M., Hogan, C. J. \& Peebles,
P. J. E., Astrophys. J. 503, 518-530 (1998)
\bibitem[7]{optical}Bernstein, R., Ph. D. Thesis, Caltech (1998)
\bibitem[8]{fir}Schlegel, D. J., Finkbeiner, D. P., \& Davis, M.,
Astrophys. J. 500, 525-553 (1998)(astro-ph/9710327)
\bibitem[9]{dirbe}Hauser, M. G., et al., Astrophys. J., 508,
25-43  (1998) (astro-ph/9806167)
\bibitem[10]{madau}Madau, P.,
in 
 Xth Rencontres de Blois meeting ``The Birth of Galaxies'', eds. B.
     Guiderdoni, F. R. Bouchet, Trinh X. Thuan,
 \& Tran Thanh Van (Gif-sur-Yvette: Edition Frontieres, 1999))(astro-ph/9812087)
\bibitem[11]{absmetals}
Pettini, M., astro-ph/9902173, in ``Chemical Evolution
from Zero to High Redshift'', ed. J. Walsh and M. Rosa (Berlin: Springer)
\bibitem[12]{clustermetals}
Renzini, A., astro-ph/9902361, ibid.
\bibitem[13]{metals}Pei, Y. C., Fall, M., \& Hauser, M. G.,
Astrophys. J., in press (1999)(astro-ph/9812182)
\bibitem[14]{qlf}Schmidt, M., Schneider, D. P. \& Gunn, J. E.,
Astron. J. 110, 68-77 (1995)
\bibitem[15]{galbh}Magorrian, J., et al., Astron. J. 115, 2285-2305 (1998)
\bibitem[16]{igm}Haardt, F. \& Madau, P., Astrophys. J. 461, 20-37 (1996)
\bibitem[17]{engine}Blandford, R. D. \& Znajek, R. L., 
Mon. Not. R. Astron. Soc. 179, 433 (1977)
\bibitem[18]{engine2}Rees, M. J., Ann. Rev. Astron. Astrophys. 22,
471-506 (1984)
\bibitem[19]{engine3}Begelman, M. C. \& Rees, M. J.,
{\it Gravity's Fatal Attraction: Black Holes in the Universe},
W. H. Freeman (1996)
\bibitem[20]{sn2}Arnett, W. D., {\it Supernovae and Nucleosynthesis:
An Investigation of the History of Matter, from the Big Bang to the Present},
Princeton (1996)
\bibitem[21]{lisaligo} See http://lisa.jpl.nasa.gov/,
http://ligo.caltech.edu/
\bibitem[22]{sn1a}Nomoto, K., Iwamoto, K. \& Kishimoto, N., Science 276,
1378-1382 (1997)
\bibitem[23]{snu1}Totani, T. \& Sato, K., Astropart. Phys. 3, 367-376 (1995)
\bibitem[24]{snu2}Hartmann, D. H. \& Woosley, S. E., Astropart. Phys.
7, 137-146 (1997)
\bibitem[25]{fireball1}Meszaros, P., \& Rees, M., Astrophys. J. 405,
278-284 (1993)
\bibitem[26]{fireball2}Meszaros, P., Rees, M. and Wijers,  R. A. M. J.,
New Astronomy, in press (1998)(astro-ph/9808106)
\bibitem[27]{distance}Mezger, M. R. et al., Nature 387, 878-880 (1997)
\bibitem[28]{energy} Kulkarni, S. R. et al., Nature, 393, 35-39 (1998);
see also
$ {\rm http://gcn.gsfc.nasa.gov/gcn/gcn3\_archive.html}$
\bibitem[29]{donut}Paczynski, B., Acta Astron. 41, 257 (1991)
\bibitem[30]{donut2}Narayan, R., Paczynski, B.,\& Piran, T.,
Astrophys. J. 395, L83-L86 (1992)
\bibitem[31]{hypernova}Paczynski, B.   Astrophys. J. 494, L45-L48 (1998)
\bibitem[32]{binaryns}Ruffert, M. \& Janka, H.- Th., Astron. Astrophys.
in press (1999)(astro-ph/9809280)
\bibitem[33]{mag}Kronberg, P. P., Rep. Prog. Phys. 57, 325-382 (1994)
\bibitem[34]{gwprojects}Abramovici, A. et al.,
Science 256, 325 (1989) 
\bibitem[35]{gw2}Thorne, K., in {\it Three Hundred Years of Gravitation}
ed. S. W. Hawking \& W. Israel, 330-446 (Cambridge, 1987)
\bibitem[36]{macho}Alcock, C. et al., Astrophys. J. 479, 119-146 (1997)
\bibitem[37]{projects}Stubbs, C. W., preprint (1998)(astro-ph/9810488)

\end{thebibliography}
\end{document}